%% file: paper_arxiv.tex
\documentclass[12pt]{article}
\include{_preamble}

\zexternaldocument*[supp:]{sm}

\title{\titlepaper}
\date{\today}
\author{\authorlist}

\begin{document}
\maketitle

\begin{abstract}
  We present an end-to-end methodological framework for causal segment discovery
  that aims to uncover differential impacts of treatments across subgroups of
  users in large-scale digital experiments. Building on recent developments in
  causal inference and non/semi-parametric statistics, our approach unifies two
  objectives: (1) the discovery of user segments that stand to benefit from
  a candidate treatment based on subgroup-specific treatment effects, and (2)
  the evaluation of causal impacts of dynamically assigning units to a study's
  treatment arm based on their predicted segment-specific benefit or harm. Our
  proposal is model-agnostic, capable of incorporating state-of-the-art machine
  learning algorithms into the estimation procedure, and is applicable in
  randomized A/B tests and quasi-experiments. An open source \texttt{R} package
  implementation, \texttt{sherlock}, is introduced.
\end{abstract}

%%%%%%%%%%%%%%%%%%%%%%%%%%%%%%%%%%%%%%%%%%%%%%%%%%%%%%%%%%%%%%%%%%%%%%%%%%%%%%%
\thispagestyle{firststyle}
\section{Introduction}\label{intro}

The ease and continued improvement in the development and deployment of digital
experiments across a variety of platforms has allowed researchers access to
cohorts larger and more diverse than previously possible. Whether in randomized
experiments or quasi-experiments (i.e., observational studies), a candidate
treatment or intervention under investigation hardly ever exhibits a uniform
effect across all units enrolled in a study. Typically, studies reveal causal
effects that differentially impact segments, or subgroups, of enrolled units,
with treatments benefiting a subset of units while posing harm to, or having no
effect upon, others --- possibly unfairly impacting sensitive groups or yielding
benefits too limited to justify the costs associated with deploying the
treatment at scale.

At Netflix, most product changes are liable to result in differential impacts
across segments of our diverse and growing global user-base of over 200M.
Whether the experimental changes are to streaming algorithms, the application’s
user interface, or outreach to users, the causal effects of treatments of
interest may vary strongly across such user characteristics as viewing habits,
the type of viewing device, household make-up, or global region, to name but
a few. Providing the best experience for all of our users and ensuring the
equitability of treatment allocation in our digital experiments is imperative to
the continuous improvement of our platform; moreover, understanding treatment
effect heterogeneity across segments of our eclectic user-base plays a critical
role in our product improvement endeavors.

Recent developments in causal inference, studying estimation of conditional
treatment effects and the population-level causal effects of dynamically
assigning treatments based on units'
characteristics~\citep{vanderweele2019selecting} have provided novel statistical
techniques that may be unified into a single framework for both discovering
population segments benefiting from a treatment~\citep{luedtke2016super} and
evaluating the causal effects of dynamic treatment rules that assign to the
treatment arm only those segments predicted to
benefit~\citep{luedtke2017evaluating}. Notably, these recent statistical
advances have been formulated in a manner capable of incorporating
state-of-the-art machine learning algorithms in the estimation procedure while
simultaneously fulfilling the theoretical requirements for valid statistical
inference on the estimated treatment effects~\citep{luedtke2016statistical}.

The remainder of this manuscript is organized as follows. In
Section~\ref{framework}, we review the foundations of a statistical framework
for identifying population segments based upon treatment effect heterogeneity
and estimating the causal effects of dynamic treatment rules assigned based upon
the conditional average treatment effect. Section~\ref{sherlock} describes
\texttt{sherlock}, a software package we designed to allow our data scientists
to perform causal machine learning for population segment discovery and
analysis. In Section~\ref{example}, we present an illustrative example of
applying our causal segment discovery framework, and the \texttt{sherlock}
package, to analyzing synthetic data inspired by quasi-experiments at Netflix.

%%%%%%%%%%%%%%%%%%%%%%%%%%%%%%%%%%%%%%%%%%%%%%%%%%%%%%%%%%%%%%%%%%%%%%%%%%%%%%%
\section{A Causal Segment Discovery Framework}\label{framework}

We consider the problem of discovering segments of users that should benefit
from a given treatment subject to an optimality criterion. More precisely, for
a dataset collected in a randomized A/B test or in a quasi-experiment, we measure
baseline covariates $W$, the assigned treatment $A$, and response/outcome $Y$ on
a set of $n$ units. We additionally assume access to a set of segmentation
covariates $V$ that temporally precede treatment assignment (and, thus, are
unaffected by it), with which we aim to segment the population. This set of
segmentation covariates may be high-dimensional (i.e., consisting of many
covariates), have high cardinality (i.e., a large support space for at least
a single covariate), or both. We define a segment as a particular realization of
these covariates (i.e., $V=v$). The segmentation covariates are often a subset
of the baseline covariates (i.e., $V \subset W$), which, in the case of
non-randomized studies, may be used to adjust for confounding of the
treatment-outcome relationship, or, in A/B tests, to reduce estimation variance
by covariate adjustment. For a given segment $V=v$, the \textit{conditional
average treatment effect} (CATE) on this segment is given by $\text{CATE}(v)
= \E[\E(Y \mid A=1,W) - \E(Y \mid A=0,W) \mid V=v]$.

The goal of segment discovery is to reveal which segments ought to benefit from
treatment, with the possible benefit being defined in two distinct ways, as
enumerated below.
\begin{enumerate}
  \item \textit{In absolute terms}, where the goal is to identify segments whose
    $\text{CATE}(v)$ estimates fall above a user-specified benefit threshold
    $\theta$, i.e., $T = \{v: \text{CATE}(v) > \theta\}$.
  \item \textit{Subject to cost or side-effect constraints}, which can be
    formulated as a constrained optimization problem with respect to the CATE.
    Specifically, the goal in this case is to identify a set of segments $T$ so
    as to maximize the population-level outcome
    \begin{equation}\label{eqn:cost}
       \psi = \sum_{v \in T} \E[\E(Y \mid A=1, W) \mid V= v] p(v) +
       \sum_{v \notin T} \E[\E(Y \mid A=0, W) \mid V= v] p(v),
    \end{equation}
subject to a cost constraint $\sum_{v \in T} \text{Cost}(v) p(v) \leq \phi$.
Here, $\text{Cost}(v) \geq 0$ is the known cost of treating a unit in a given
segment, the budget $\phi$ is a given spending cap, and $p(v)$ is the marginal
distribution of segment $v$ in the population. This is equivalent to the binary
knapsack problem of finding $T \subseteq \{v: \text{CATE}(v) > 0\}$ such that we
maximize $\sum_{v \in T} \text{CATE}(v)$ while respecting the constraint
$\sum_{v \in T} \text{Cost}(v) p(v) \phi$.
\end{enumerate}

Estimation of the CATE requires the use of sample splitting principles in order
to avoid
overfitting that stems from learning the dynamic treatment rule and evaluating
the population-level causal effects of the learned rule within a single study
cohort~\citep{vdl2015targeted}. Similar principles were studied in the
formulation of cross-validated targeted minimum loss estimation
(CV-TMLE)~\citep{zheng2011cross}, have recently been renamed
``cross-fitting''~\citep{chernozhukov2017double}, and appeared decades ago in
studies of efficient estimation in semiparametric
models~\citep{klaassen1987consistent, bickel1993efficient}. Following the
partitioning of the dataset into $K$ training and validation folds (i.e.,
$K$-fold cross-validation), we apply user-specified machine learning algorithms
to obtain out-of-sample (or holdout) predictions of the outcome regression
$\hat{\E}_Y (Y \mid A,W)$, the conditional mean of the outcome given observed
treatment status and covariates, and of the propensity score $\hat{p}_A (A \mid
W)$, the conditional probability of treatment assignment given covariates. Next,
using the form of the efficient influence function~\citep{bickel1993efficient},
a central quantity in semiparametric theory, a doubly robust outcome
transformation may be applied to obtain the pseudo-outcome $\hat{D} = [(2A - 1)
/ \hat{\prob}_A (A \mid W)] \cdot [Y - \hat{\E}_Y(Y \mid A, W)] + \hat{\E}_Y(Y
\mid 1, W) - \hat{\E}_Y(Y \mid 0, W)$. Regressing the outcome $\hat{D}$ on
segmentation covariates $V$, by way of another user-specified machine learning
algorithm, yields $\hat{\E}_V(\hat{D} \mid V)$, a \textit{doubly robust}
estimator of $\text{CATE}(v)$, which admits analytical standard error
calculations based on its efficient influence function. For both problems (1)
and (2), a single-tailed test of the null and alternative hypotheses $\{H_0:
\text{CATE}(v) \leq \theta, H_1: \text{CATE}(v) > \theta\}$, across the segments
can be used to assign the treatment rule inferentially, with multiple
comparisons corrections to account for the number of segments. For problem (2),
only segments for which $\text{CATE}(v) > 0$ (i.e., those expected to benefit
from treatment) are considered as candidates in the binary knapsack problem.

After learning the segments that should receive treatment, we can assess the
gain attributable to such dynamic treatment strategies, by comparison against
a global (or ``one-size-fits-all'') treatment strategy. A dynamic treatment
strategy learned in this way is defined as a function $A=d(v)$, where $d(v)$ is
the indicator that $v \in T$, that is, whether a given user belongs to
a ``should-treat'' segment. At this point, doubly robust estimators of optimal
treatment effects (OTEs), such as $\psi_{\text{OTE}} = \E[\E(Y \mid A = d(V),
W) - \E(Y \mid A=1, W)]$, or the heterogeneous treatment effect (HTE) over the
segments, namely, $\psi_{\text{HTE}} = \E[\E(Y \mid A=1, W) - \E(Y \mid A=0,W)
\mid V \in T] - \E[\E(Y \mid A=1, W) - \E(Y \mid A=0, W) \mid V \notin T]$, may
be used to evaluate the quality of the population segmentation based on the
expected utility of deploying the resultant dynamic treatment strategy.

%%%%%%%%%%%%%%%%%%%%%%%%%%%%%%%%%%%%%%%%%%%%%%%%%%%%%%%%%%%%%%%%%%%%%%%%%%%%%%%
\section{\texttt{sherlock}: A Modular and Scalable Software
  Implementation}\label{sherlock}

We developed \texttt{sherlock}, a software package for the \texttt{R}
programming language and environment for statistical computing~\citep{R}, to
simplify the use of our causal segmentation framework in both randomized A/B
testing and quasi-experimental study settings. The design of this \texttt{R}
package emphasized two key goals: \textit{modularity} and \textit{scalability}.
In terms of modularity, the package abstracts the causal segmentation analysis
pipeline into three broad steps, each representing a distinct point of
analytical interest and encoded in one of \texttt{sherlock} user-facing wrapping
functions.
\begin{enumerate}
  \item Evaluate treatment effect heterogeneity across the segment groups,
     requiring estimation of nuisance parameters as well as estimation of the
     CATE on the population segments $v \in V$. This step involves computing
     estimates of the propensity score, the outcome regression, and of
     $\text{CATE}(v)$, for which a user-specified modeling algorithm (or
     algorithm library) is used. In particular, we recommend the Super Learner
     algorithm, which allows the selection of a single algorithm (or the
     construction of an ensemble model) from a user-specified algorithm library,
     using loss-based cross-validation to achieve optimal asymptotic
     performance~\citep{vdl2007super}. To facilitate access to a wide variety of
     modeling algorithms, \texttt{sherlock} is tightly coupled with the open
     source \texttt{sl3} \texttt{R} package~\citep{coyle2021sl3}, which acts
     both as an interface to 50+ machine learning algorithms and a standalone
     implementation of the Super Learner algorithm. This step is encoded in
     the \texttt{sherlock\_calculate()} function.
  \item Dynamically assign treatment to (or withhold treatment from) the
    population segments, based on options for evaluating treatment efficacy,
    including thresholding the CATE at a user-given cutoff or while accounting
    for an overall budget or cost for deploying the
    treatment~\citep{luedtke2016optimal, vanderweele2019selecting}. This step
    uses the estimated $\text{CATE}(v)$ and user-specified optimality criterion
    (e.g., total budget) to assign segments with $\text{CATE}(v) > 0$ to the
    treatment arm, constructing confidence intervals and performing hypothesis
    tests (i.e., $H_0: \text{CATE}(v) \leq \theta, H_1: \text{CATE}(v)
    > \theta$). The data scientist may now examine a summary of the segmentation
    results (see Table~\ref{tab:sherlock_results}), which includes segment
    identifiers $v \in V$, point estimates of $\text{CATE}(v)$, inference
    for the estimates, and an indication of a segment's treatment assignment
    (i.e., $v \in T$). This is implemented by the \texttt{watson\_segment()}
    function.
  \item Estimate population-level causal effects of the dynamic treatment rule,
    including both HTEs, which contrast the effects of the intervention within
    the treatment allocations derived in the preceding step, and OTEs, which
    contrast counterfactual means under the dynamic treatment rule with those of
    static treatment strategies. This step allows for the evaluation of
    segmentation quality through the use of non/semi-parametric efficient
    estimation of the effect measures discussed in Section~\ref{framework}.
    Notably, neither this step nor the preceding step is computationally
    demanding, relying primarily on the nuisance estimates generated in the
    first step. These effects are computed by the \texttt{mycroft\_assess()}
    function.
\end{enumerate}

With respect to scalability, \texttt{sherlock}'s design emphasized enumerating
the set of nuisance functionals required for doubly robust estimation of the
CATE and the causal effect measure parameters (i.e., HTE, OTE). By accounting
for dependencies across stages of the estimation procedure at the design stage,
\texttt{sherlock}'s pipeline collapses computationally taxing nuisance
regressions into a single step, thereby streamlining the latter steps, in which
the segment-specific treatment rule is assigned and the quality of segmentation
evaluated. This reduced computational burden is a direct result of (1) avoiding
redundant computation and (2) breaking dependencies between steps of the
estimation routine to introduce modularization. Together, these design choices
allow the data science practitioner increased flexibility when using
\texttt{sherlock} to perform causal segmentation analysis, allowing for choices
of the treatment rule optimality criterion and effect measure to be siloed
entirely from nuisance estimation --- that is, the nuisance estimate ``building
blocks'' may easily be extracted and manipulated, without any need for
re-computation, as the analytical decision-making process progresses.

Finally, \texttt{sherlock} includes (by way of \texttt{sl3}) facilities to
introduce customizable wrappers that accommodate newly developed machine
learning algorithms or regression techniques for the estimation of nuisance
parameters. In our application settings --- working with large-scale experiments
with millions of users --- this proves an important point of flexibility, as
regression algorithms do not uniformly scale to massive sample sizes. Instead,
the analyst may, with little overhead, build bindings for algorithms optimized
to the scale and characteristics of the dataset at hand, including, for example,
algorithms optimized to respect measurement sparsity. To support
cross-validation of nuisance estimates, \texttt{sherlock} relies on the
\texttt{origami} \texttt{R} package~\citep{coyle2018origami}, which provides
a general and flexible framework for sample splitting. Both \texttt{origami} and
\texttt{sl3} utilize the \texttt{Future} framework~\citep{bengtsson2021unifying}
for delayed and parallelized computation in \texttt{R}, allowing
\texttt{sherlock} to take full advantage of available compute resources;
moreover, \texttt{sherlock}'s data structures are simple instantiations of
\texttt{R}'s \texttt{data.table}~\citep{dowle2021dt}, which provides numerous
optimizations that streamline computation, including reference assignment
operations that curb memory usage, helpful when working with massive datasets.
An open source release of our \texttt{sherlock} \texttt{R} package is available
on GitHub at \url{https://github.com/Netflix/sherlock}, and, in time,
\texttt{sherlock} will additionally be made available via the Comprehensive
\texttt{R} Archive Network (CRAN).

%%%%%%%%%%%%%%%%%%%%%%%%%%%%%%%%%%%%%%%%%%%%%%%%%%%%%%%%%%%%%%%%%%%%%%%%%%%%%%%
\section{Illustrative Example}\label{example}

\begin{figure}[H]
  \centering
  \includegraphics[scale=0.28]{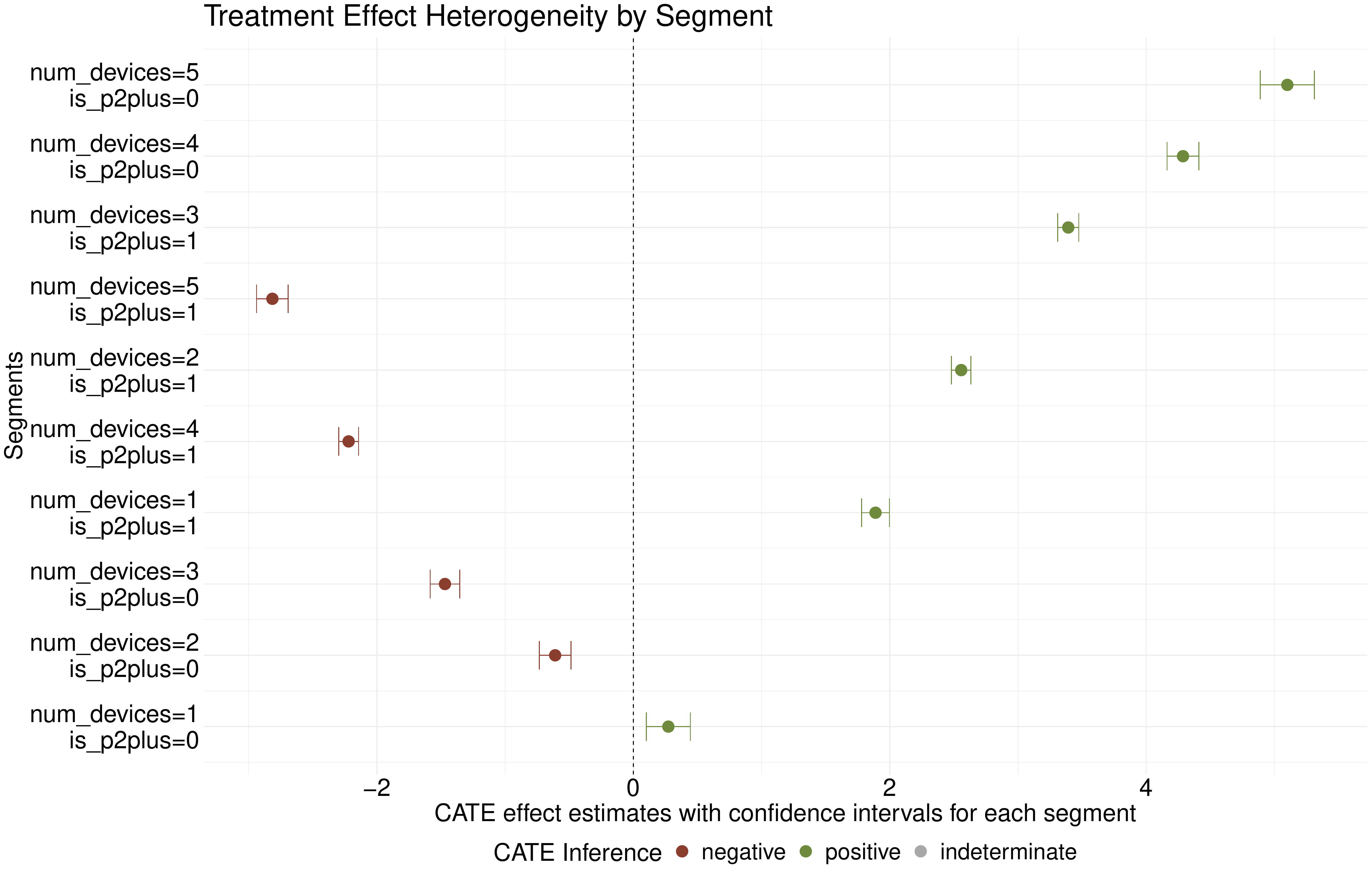}
  \caption{Results of causal segmentation analysis, with statistical inference
    for the CATE. Deploying the new UI is expected to lower the viewing hours
    of segments with negative CATE estimates, while yielding increases in
    viewing hours for those segments with positive estimates.}
  \label{fig:sherlock}
\end{figure}

We demonstrate the use of our causal segment discovery framework by the
application of \texttt{sherlock} to a synthetic dataset example inspired by
quasi-experiments at Netflix. In our simulated dataset, we have $n = 5000$
independent units (e.g., Netflix accounts), upon which we have measured $7$
variables, including the number of devices associated with a given account
(``num\_devices''), whether the account belongs to a newly enrolled Netflix
member (``is\_p2plus''), whether the account is located in a new market region
(``is\_newmarket''), the account's estimated lifetime value (``baseline\_ltv''),
a measure of the account's baseline viewing hours (``baseline\_viewing''),
non-randomized assignment to a new user interface (UI; the treatment), and
a measure of post-treatment Netflix viewing hours (``outcome\_viewing''). A few
randomly selected rows of the dataset appear in Table~\ref{tab:example_summary}
for reference.
\vspace{-1em}
\input{tables/data_summary}
\vspace{-2em}
The candidate segmentation covariates are the number of devices associated with
the account and whether the account belongs to a new member. Across these two
segmentation covariates, we observe $10$ strata, that is, $\text{card}(V) = 10$.
Next, we evaluate treatment effect heterogeneity by estimating the CATE, using
a handful of regression and machine learning algorithms to fit nuisance
parameters. By calling the function \texttt{sherlock\_calculate()} and its
associated plotting method, we can easily visualize the CATE estimates across
each segment; these are displayed in Figure~\ref{fig:sherlock}.
\vspace{-1em}
With CATE estimates for each population segment, we are now prepared to decide
which subgroups ought to have the new UI experience rolled out to them. Using
\texttt{sherlock}'s \texttt{watson\_segment()} function, we can evaluate the
evidence for a segment benefiting from treatment, thresholding the CATE at its
null value of $\theta = 0$ (i.e., no treatment effect) and correcting for
hypothesis testing multiplicity appropriately; the results are summarized in
Table~\ref{tab:sherlock_results}.
\vspace{-1em}
\input{tables/sherlock_summary}
\vspace{-2em}

%%%%%%%%%%%%%%%%%%%%%%%%%%%%%%%%%%%%%%%%%%%%%%%%%%%%%%%%%%%%%%%%%%%%%%%%%%%%%%%
\subsection*{Acknowledgements}

\footnotesize{
We thank Simon Ejdemyr and Martin Tingley, for helpful discussions and feedback
on an early draft of this manuscript; Stephanie Lane, for helpful guidance on
\texttt{sherlock}'s native data visualization capabilities; and Mark van der
Laan, for insightful remarks on the implementation of previously described
algorithms for targeted minimum loss estimation.
}

%%%%%%%%%%%%%%%%%%%%%%%%%%%%%%%%%%%%%%%%%%%%%%%%%%%%%%%%%%%%%%%%%%%%%%%%%%%%%%%
\footnotesize{\bibliography{refs}}
\end{document}

%% file: _preamble.tex
\usepackage{nameref,zref-xr} % nameref before zref-xr
\zxrsetup{toltxlabel} % allows use of the \ref command as usual
\usepackage{url}
\usepackage[colorlinks,citecolor=blue,urlcolor=blue]{hyperref}

\usepackage{arxiv}
\usepackage[inline,shortlabels]{enumitem}
\usepackage[english]{babel}
\usepackage{longtable}
\usepackage{color}
\usepackage{multirow}
\usepackage{setspace}
\usepackage{dsfont}
\usepackage[OT1]{fontenc}
\usepackage[utf8]{inputenc}
\usepackage{lastpage}
\usepackage{standalone}
\usepackage{booktabs}
\usepackage{refcount}
\usepackage{subcaption}
\usepackage{float}
\usepackage{graphicx}
\graphicspath{ {figs/} }

\usepackage{amsmath,amsfonts,amsthm}
\usepackage{bbm,bm,mathtools}
\usepackage[linesnumbered,ruled,vlined]{algorithm2e}

\usepackage[numbers]{natbib}
 \newcommand{\titlepaper}{A Framework for Causal Segmentation Analysis with
   Machine Learning in Large-Scale Digital Experiments}

\newcommand{\authorlist}{
  Nima S.~Hejazi \\
  Data Science \& Engineering\\
    Netflix\\
  \texttt{nhejazi@berkeley.edu} \\
  \And
  Wenjing Zheng \\
  Data Science \& Engineering\\
    Netflix\\
  \texttt{wzheng@netflix.com} \\
  \And
  Sathya Anand \\
  Data Science \& Engineering\\
    Netflix\\
  \texttt{sathya@netflix.com} \\
}

\fancypagestyle{firststyle}{
 \fancyhf{}
 \cfoot{\footnotesize \textit{ Manuscript accepted by the
   8\textsuperscript{th} annual Conference on Digital Experimentation (CODE)
   @ MIT}}
}
\newcommand{\E}{\mathbb{E}}

\newcommand{\prob}{\mathbb{P}}

\AtEndDocument{\refstepcounter{theorem}\label{finalthm}}
\AtEndDocument{\refstepcounter{equation}\label{finaleq}}
{\theoremstyle{definition} }

{\theoremstyle{definition}}

\AtEndDocument{\refstepcounter{lemma}\label{finallemma}}

%% file: tables/data_summary.tex
\begin{table}[H]
\centering
\resizebox{0.88\columnwidth}{!}{%
\begin{tabular}{r|r|r|r|r|r|r}
\hline
num\_devices & is\_p2plus & is\_newmarket & baseline\_ltv & baseline\_viewing &
treatment & outcome\_viewing\\
\hline
3 & 0 & 1 & 0.539 & 0.000 & 1 & 0.406\\
\hline
2 & 1 & 1 & 1.328 & 1.637 & 0 & 2.328\\
\hline
3 & 1 & 0 & 0.000 & 0.000 & 1 & 3.400\\
\hline
2 & 1 & 0 & 1.027 & 0.000 & 0 & 1.934\\
\hline
2 & 1 & 1 & 0.000 & 0.000 & 0 & 1.376\\
\hline
3 & 0 & 1 & 0.000 & 1.401 & 0 & 2.683\\
\hline
\end{tabular}%
}
\vspace{0.3em}
\caption{
  Measurements collected on six units in a simulated quasi-experimental
  dataset. Assessing the impact of the new UI, itself possibly affected by the
  baseline covariates, on the outcome of viewing hours is the question of
  interest.
}
\label{tab:example_summary}
\end{table}

%% file: tables/sherlock_summary.tex
\begin{table}[H]
\centering
\resizebox{0.88\columnwidth}{!}{%
\begin{tabular}{r|r|r|r|r|r|r|r|r}
\hline
num\_devices & is\_p2plus & Segment Proportion & CATE & Lower CL & Upper CL & Std.~Err.~& p-value & Treat?\\
\hline
1 & 0 & 0.0392 & 0.283 & 0.108 & 0.459 & 0.089 & 0.001 & Yes\\
\hline
1 & 1 & 0.0862 & 1.895 & 1.787 & 2.00 & 0.055 & 0.000 & Yes\\
\hline
2 & 0 & 0.0718 & -0.595 & -0.719 & -0.471 & 0.063 & 1.000 & No\\
\hline
2 & 1 & 0.1752 & 2.550 & 2.473 & 2.626 & 0.039 & 0.000 & Yes\\
\hline
3 & 0 & 0.0764 & -1.462 & -1.580 & -1.343 & 0.061 & 1.000 & No\\
\hline
3 & 1 & 0.1796 & 3.391 & 3.308 & 3.473 & 0.042 & 0.000 & Yes\\
\hline
4 & 0 & 0.0728 & 4.304 & 4.180 & 4.428 & 0.063 & 0.000 & Yes\\
\hline
4 & 1 & 0.1814 & -2.214 & -2.291 & -2.137 & 0.039 & 1.000 & No\\
\hline
5 & 0 & 0.0328 & 5.094 & 4.879 & 5.310 & 0.119 & 0.000 & Yes\\
\hline
5 & 1 & 0.0846 & -2.798 & -2.921 & -2.674 & 0.063 & 1.000 & No\\
\hline
\end{tabular}%
}
\vspace{0.3em}
\caption{Point estimates of $\text{CATE}(v)$, with lower and upper confidence
limits across the ten segments, as well as p-values for hypothesis tests of
$H_0: \text{CATE}(v) \leq \theta, H_1: \text{CATE}(v) > \theta$ and
corresponding treatment decisions.}
\label{tab:sherlock_results}
\end{table}